\documentclass{emulateapj}

\def\nul#1{}
\def \vv#1{\mathbf{#1}}
\def \be {\begin{equation}}
\def \ee {\end{equation}}
\def \k  {s}
\def \K  {k}
\def \L  {0}
\def \p  {0}
\def \s  {1}
\def \S  {s}
\def \ij {i}
\def \ji {j}
\def \eq {{}}
\def \es {f}
\def \de {\theta}

\def\ve{\varepsilon}

\def\Frac#1#2{{{\displaystyle\strut#1}\over{\displaystyle\strut#2}}}

\def\crm{\cr\noalign{\medskip}}

\def \llabel#1{\label{#1}}

\def\apjl{Astrophys.\ J. }
\def\apjs{Astrophys.\ J.\ (Supp.) }
\def\aap{Astron.\ Astrophys. }

\shorttitle{Secular evolution of a satellite by tidal effect}
\shortauthors{A.C.M. Correia}

\begin{document}


\title{Secular evolution of a satellite by tidal effect. Application to Triton.}


\author{Alexandre C.M. Correia}
\affil{Departamento de F\'\i sica, Universidade de Aveiro, Campus de
Santiago, 3810-193 Aveiro, Portugal. \\
Astronomie et Syst\`emes Dynamiques, IMCCE-CNRS UMR8028, 
77 Av. Denfert-Rochereau, 75014 Paris, France}

\email{correia@ua.pt}

\begin{abstract}
Some of the satellites in the Solar System, including the Moon, appear to
have been captured from heliocentric orbits at some point in their past, 
and then have evolved to the present configurations.
The exact process of how this trapping occurred is unknown, but the dissociation
of a planetesimal binary in the gravitational field of the planet, gas drag, or a
massive collision seem to be the best candidates. 
However, all these mechanisms leave the satellites in elliptical orbits
that need to be damped to the present almost circular ones. 
Here we give a complete description of the secular tidal evolution of a satellite
just after entering a bounding state with the planet.
In particular, we take into account the spin evolution of the satellite, which
has often been assumed synchronous in previous studies. 
We apply our model to Triton and successfully explain some geophysical
properties of this satellite, as well as
the main dynamical features observed for the Neptunian system.
\end{abstract}



\keywords{celestial mechanics ---
methods: analytical ---
planets and satellites: general ---
planets and satellites: individual (Neptune, Triton)}


\section{Introduction}

Both the Earth's Moon and Pluto's moon, Charon, have an important fraction of the mass
of their systems, and therefore they could be classified as double-planets rather than
as satellites. The proto-planetary disk is unlikely to produce such systems, and their
origin seems to be due to a catastrophic impact of the initial planet with a body of
comparable dimensions \cite[e.g.][]{Canup_Asphaug_2001,Canup_2005}.
On the other hand,
Neptune's moon, Triton, and the Martian moon, Phobos, are spiraling down into the
planet, clearly indicating that the present orbits are not primordial, and may
have undergone a long evolving process from a previous capture
\citep[e.g.][]{Mignard_1981m,Goldreich_etal_1989}.

The present orbits of all these satellites are almost circular, and their spins
appear to be synchronous with the orbital mean motion, as well as being locked in Cassini
states \citep[e.g.][]{Colombo_1966,Peale_1969}.
This also applies to the Galilean satellites of Jupiter, which are likely to
have originated from Jupiter's accretion disk and additionally show
orbital mean motion resonances \citep[e.g.][]{Yoder_1979}.
All these features seem to be due to tidal evolution, which arises
from differential and inelastic deformation of the planet by a perturbing body.

Previous long-term studies on the orbital evolution of satellites 
have assumed that their rotation is synchronously locked, and therefore
limits the tidal evolution to the orbits \citep[e.g.][]{McCord_1966}.
However, these two kinds of evolution cannot be dissociated because the total
angular momentum must be conserved. 
Additionally, it has been assumed that the spin axis is locked in a Cassini
state with very low obliquity. Although these assumptions are correct for the
presently known situations, they were not necessarily true throughout the evolution.

The aim of this Letter is to model the orbital evolution of a satellite from
its origin or capture until the preset day, including spin
evolution for both planet and satellite, and also to make predictions regarding
its future evolution.
We provide a simple averaged model 
adapted for fast computational simulations, as required for long-term studies.
In the last section we apply this model to the Triton-Neptune system.

\section{The model}

We consider a hierarchical system composed of a star, a planet
and a satellite, with masses $M \gg m_\p \gg m_\s$, respectively.
Both planet and satellite are considered oblate ellipsoids with gravity field
coefficients given by $J_{2_\p}$ and $J_{2_\s}$, rotating about the axis of maximal
inertia along the directions $\vv{\hat \k}_\p$ and $\vv{\hat \k}_\s$, with rotation
rates $\omega_\p$ and $\omega_\s$, respectively.
The potential energy $U$ of the system is then given by 
\citep[e.g.][]{Smart_1953}: 

\begin{eqnarray}
U & =&  
- G \frac{M m_\p}{r_\p} \left( 1 - \sum_{i=0,1} J_{2_\ij} \frac{m_\ij}{m_\p}
\left(\frac{R_\ij}{r_\p}\right)^2 \!\! P_2 (\vv{\hat r}_\p \cdot \vv{\hat \k}_\ij)
\right) \crm && 
- G \frac{m_\p m_\s}{r_\s} \left( 1 - \sum_{\ij=0,1} J_{2_\ij}
 \left(\frac{R_\ij}{r_\s}\right)^2 \!\! P_2 (\vv{\hat r}_\s \cdot \vv{\hat \k}_\ij) 
\right) \crm &&  
- G \frac{M m_\s}{r_\p} \left(\frac{r_\s}{r_\p}\right)^2 \!\! P_2 (\vv{\hat
r}_\p \cdot \vv{\hat r}_\s) \ , \llabel{090514a}
\end{eqnarray}
where terms in $(R_\ij/r_\ji)^3$ have been neglected
($\ij,j=0,1$). $G$ is the gravitational constant, $R_\ij$ the radius of the planet
or the satellite, $\vv{r}_\ij$ the distance between the planet and the star or the
satellite, and $P_2(x) = (3x^2-1)/2$ the Legendre polynomial of degree two.

Neglecting tidal interactions with the star, the tidal potential is written
\citep[e.g.][]{Kaula_1964}:
\be
U_T = - G \frac{m_\p m_\s}{r_\s^3} \sum_{\ij=0,1} k_{2_\ij}
\frac{R_\ij^5}{r_\ij'^3} P_2 (\vv{\hat r}_\s \cdot \vv{\hat r}_\ij')  \ ,
\llabel{090514b} 
\ee
where $k_{2_\ij}$ is the potential Love number for the planet or the
satellite, and $\vv{r}_\ij'$ the position of the interacting body
at a time delayed of $\Delta t_\ij$. For simplicity, we will adopt a model with
constant $\Delta t_\ij$, which can be made linear 
\citep[e.g.][]{Mignard_1979,Hut_1981,Surgy_Laskar_1997}:
\be
\vv{r}_\ij' \simeq \vv{r}_\s + \Delta t_\ij \left(\omega_\ij \vv{\k}_\ij
\times \vv{r}_\s - \vv{\dot r}_\s \right) \ . \llabel{090514c}
\ee

The complete evolution of the system can be tracked by the evolution of the
rotational angular momentums, $ \vv{H}_\ij \simeq C_\ij \omega_\ij \vv{\hat
\k}_\ij $, the orbital angular momentums, $\vv{L}_\ij \simeq m_\ij n_\ij a_\ij^2
(1-e_\ij^2)^{1/2} \vv{\hat \K}_\ij $, and the orbital energy $E_1 \simeq - G m_\p m_\s /
(2 a_\s) $. 
$n_\ij$ is the mean motion, $a_\ij$ the semi-major axis, $e_\ij$ the
eccentricity, and $C_\ij$ the principal moment of inertia. The contributions
to the orbits are easily computed from the above potentials as 
\be
\vv{\dot L}_\p = - \vv{r}_\p \times \vv{\nabla}_{\!\vv{r}_\p} U ; \quad
\vv{\dot L}_\s = - \vv{r}_\s \times \vv{\nabla}_{\!\vv{r}_\s} U ; \quad
\dot{E}_\s = - \vv{\dot r}_\s \cdot \vv{\nabla}_{\!\vv{r}_\s} U  \ .
\llabel{090514d}
\ee

Since the total angular momentum is conserved, the contributions to the spin of
the planet and satellite can easily be computed from the orbital contributions: 
\be 
\vv{\dot H}_\p + \vv{\dot H}_\s + \vv{\dot L}_\p + \vv{\dot L}_\s = 0 \ .
\llabel{090514e}
\ee

Because tidal effects act in long-term time-scales, we further average the
equations of motion over fast angles, namely the true anomaly and the longitude
of the periapse. The resulting equations for the conservative motion
are \citep[e.g.][]{Boue_Laskar_2006}:
\begin{equation}
\vv{\dot L}_\s = - \gamma \cos I \, \vv{\hat \K}_\p \times \vv{\hat \K}_\s 
- \sum_\ij \beta_\ij \cos \de_\ij \, \vv{\hat \k}_\ij \times \vv{\hat \K}_\s 
 \ , \llabel{090514z}
\end{equation}
\begin{equation}
\vv{\dot H}_\ij = - \alpha_\ij \cos \ve_\ij \, \vv{\hat \K}_\p \times \vv{\hat 
\k}_\ij  - \beta_\ij \cos \de_\ij \, \vv{\hat \K}_\s \times \vv{\hat \k}_\ij \ ,
\llabel{090514f}
\end{equation}
where
\be 
\alpha_\ij = \frac{3 G M m_\ij J_{2_\ij} R_\ij^2}{2 a_\p^3 (1-e_\p^2)^{3/2}} \ ,
\llabel{090514g}
\ee
\be 
\beta_\ij =  \frac{3 G m_\p m_\s J_{2_\ij} R_\ij^2}{2 a_\s^3 (1-e_\s^2)^{3/2}} \ ,
\llabel{090514h}
\ee
\be 
\gamma = \frac{3 G M m_\s a_\s^2 (2+3e_\s^2)}{8 a_\p^3 (1-e_\p^2)^{3/2}} \ ,
\llabel{090514i}
\ee
and
\be
\cos \ve_\ij = \vv{\hat \k}_\ij \cdot \vv{\hat \K}_\p \ , \quad
\cos \de_\ij = \vv{\hat \k}_\ij \cdot \vv{\hat \K}_\s \ , \quad
\cos I   = \vv{\hat \K}_\p \cdot \vv{\hat \K}_\s \ , \llabel{090514j}
\ee
are the direction cosines of the spins and orbits: $\ve_\ij$ is the
obliquity to the orbital plane of the planet, $ \de_\ij$ the obliquity to the
orbital plane of the satellite, and $I$ the inclination between orbital planes.

For the dissipative tidal effects, we can obtain the equations of motion
directly from equation (\ref{090514d}), using $U_T$ instead of $U$, that is, 
$ \vv{\dot L}_\p = 0$,
\be
\vv{\dot H}_\ij = - K_\ij \,
n_\s \left( f_1(e_\s) \Frac{\vv{\hat \k}_\ij + \cos \de_\ij \,
\vv{\hat \K}_\s}{2} \Frac{\omega_\ij}{n_\s} - f_2(e_\s) \vv{\hat \K}_\s 
\right) \ , \llabel{090514k}
\ee
\be
\dot E_\s =  \sum_{\ij \ne \ji} K_\ij \, n_\s^2
\left( f_2(e_\s) \cos \de_\ij \Frac{\omega_\ij}{n_\s} - f_3(e_\s)
\right) \ , \llabel{090514l}
\ee
where ($ \ij, \ji = 0,1 $),
\be
K_\ij = \Delta t_\ij (3 k_{2_\ij} G m_\ji^2 R_\ij^5)/a_\s^6 \ , \llabel{090514m}
\ee
and
$
f_1(e) = (1 + 3e^2 + 3e^4/8)/(1-e^2)^{9/2} \ , \llabel{090514n}
$
$
f_2(e) = (1 + 15e^2/2 + 45e^4/8 + 5e^6/16)/(1-e^2)^{6} \ , \llabel{090514o}
$
$
f_3(e) = (1 + 31e^2/2 + 255e^4/8 + 185e^6/16 + 25e^8/64)/(1-e^2)^{15/2} \ . 
\llabel{090514p}
$

\section{Secular evolution}

In the previous section we presented the equations that rule the tidal
evolution of a satellite in terms of angular momenta and orbital energy.
However, the spin and orbital quantities are better represented by the 
rotation angles and elliptical elements.
The direction cosines (Eq.\ref{090514j}) are obtained from the
angular momenta vectors, since $ \vv{\hat \k}_\ij = \vv{H}_\ij / || \vv{H}_\ij
|| $ and $ \vv{\hat \K}_\ij = \vv{L}_\ij / || \vv{L}_\ij || $, as well as the
rotation rate $ \omega_\ij = \vv{H}_\ij \cdot \vv{\hat \k}_\ij / C_\ij $.
The semi-major axis and the eccentricity can be obtained from
$E_1$ and $|| \vv{L}_\s ||$, respectively.

The variation in the satellite's rotation rate can be computed from
equation (\ref{090514k}) as $\dot \omega_\ij = \vv{\dot H}_\ij \cdot \vv{\hat
\k}_\ij / C_\ij $, giving \citep{Correia_Laskar_2009}:
\be
\dot \omega_\s = - \frac{K_\s \, n_\s}{C_\s} 
\left( f_1(e_\s) \Frac{1 + \cos^2 \de_\s}{2} \Frac{\omega_\s}{n_\s} -
f_2(e_\s) \cos \de_\s \right) \ . \llabel{090515a}
\ee
For a given obliquity and eccentricity, the equilibrium rotation rate, obtained
when $ \dot \omega_\s = 0 $, is attained for:
\be
\frac{\omega_\s^\eq}{n_\s} = \frac{f_2(e_\s)}{f_1(e_\s)} \, \frac{2 \cos
\de_\s}{1 + \cos^2 \de_\s} \ , \llabel{090520a}
\ee

The obliquity variations can be obtained from equation (\ref{090514j}):
\be
\frac{d \cos \de_\ij}{d t} = \frac{\dot \vv{H}_\ij \cdot ( \vv{\hat
\K}_\s - \cos \de_\ij \vv{\hat \k}_\ij)}{|| \vv{H}_\ij ||} + \frac{\dot
\vv{L}_\s \cdot ( \vv{\hat \k}_\ij - \cos \de_\ij \vv{\hat \K}_\s)}{||
\vv{L}_\s ||}  \ . \llabel{090520b}
\ee
For the conservative motion (Eqs.\,\ref{090514z}, \ref{090514f}), 
stable configurations for the spin can be found whenever the vectors
($\vv{\hat \k}_\s$, $\vv{\hat \K}_\s$, $\vv{\hat \K}_\p$) or ($\vv{\hat \k}_\s$,
$\vv{\hat \K}_\s$, $\vv{\hat \k}_\p$) are coplanar and precess at the same rate
$ g $ \citep[e.g.][]{Colombo_1966,Peale_1969}.
The first situation occurs if $ \gamma \gg \beta_\p $ (outer satellite) and
the second situation when $ \gamma \ll \beta_\p $ (inner satellite).
The equilibrium obliquities can be found from a single relationship
\citep[e.g.][]{Ward_Hamilton_2004}: 
\be
\lambda_\s \cos \de_\s^\eq  \sin \de_\s^\eq  + \sin (\de_\s^\eq  - I_\L) = 0 \ ,
\llabel{090519b} 
\ee
where $ \lambda_\s = \beta_\s / (C_\s \omega_\s g) $ is a dimensionless parameter and
$ I_\L $ is the inclination of the orbit of the satellite with respect to
the Laplacian plane ($ I_\L \simeq I $  and $ g \simeq \gamma \cos I / || \vv{L}_\s || $
for an outer satellite, and $ I_\L \simeq \de_\p $ and $ g \simeq \beta_\p
\cos \de_\p / || \vv{L}_\s || $ for an inner satellite)
\citep[e.g.][]{Laplace_1799,Mignard_1981m,Tremaine_etal_2009}. 
The above equation has two or four real roots for $\de_\s$, which are known by $Cassini$
$states$. In general, for satellites we have $ I_\L \sim 0$, and these solutions
are approximately given by: 
\be
\tan^{-1} \left(\frac{\sin I_\L}{\cos I_\L \pm \lambda_\s}\right) \ , \quad
\pm \cos^{-1} \left(-\frac{\cos I_\L}{\lambda_\s}\right)  \ . \llabel{090519e}
\ee
For a generic value of $I_\L$, when $\lambda_\s \ll 1 $, which
is often the case of an outer satellite, the first expression gives the only two
real roots of equation (\ref{090519b}), one for $ \de_\s^\eq  \simeq I_\L $ and
another for $ \de_\s^\eq \simeq \pi - I_\L $. On the other hand, when $\lambda_\s \gg
1 $, which is the case of inner satellites, we will have four real roots
approximately given by expressions (\ref{090519e}).

In turn, the dissipative obliquity variations are computed by substituting
equation (\ref{090514k}) in (\ref{090520b}) with $ || \vv{H}_\s || \ll ||
\vv{L}_\s ||$, giving: 
\be
\dot \de_\s \simeq \frac{K_\s n_\s}{C_\s \omega_\s} \sin \de_\s
\left( f_1(e_\s) \cos \de_\s \frac{\omega_\s}{2 n_\s} - f_2(e_\s) \right) 
\ . \llabel{090520d}
\ee
Because of the factor $ n_\s / \omega_\s $ in the magnitude of the obliquity
variations, for an initial fast rotating satellite, the 
time-scale for the obliquity evolution will be longer than the time-scale
for the rotation rate evolution (Eq.\ref{090515a}). 
As a consequence, it is to be expected that the rotation rate reaches its
equilibrium value (Eq.\ref{090520a}) earlier than the obliquity.
Thus, replacing equation (\ref{090520a}) in (\ref{090520d}), we have:
\be
\dot \de_\s \simeq - \frac{K_\s n_\s}{C_\s \omega_\s} f_2(e_\s)
\frac{\sin \de_\s}{1+\cos^2 \de_\s} \ .
\llabel{090520e}
\ee
We then conclude that the obliquity can only decrease by tidal effect, since $
\dot \de_\s \le 0 $, and the final obliquity tends to be captured
in low obliquity Cassini states.



The variations in the semi-major axis are obtained from the energy variations 
$ \dot E_\s $,
\be
\dot a_\s 
=  \sum_{\ij \ne \ji} \frac{2 K_\ij}{m_\s a_\s} \,
\left( f_2(e_\s) \cos \de_\ij \frac{\omega_\ij}{n_\s} - f_3(e_\s)
\right) \ , \llabel{090515b}
\ee
while the eccentricity is obtained from the norm of the orbital angular momentum $ ||
\vv{L}_\s || = m_\s n_\s a_\s^2 (1-e_\s^2)^{1/2} $: 
\begin{eqnarray}
\dot e_\s 
& = & \sum_{\ij \ne \ji} \frac{9 K_\ij}{m_\s a_\s^2} \left( \frac{11}{18} f_4(e_\s) \cos \de_\ij
\frac{\omega_\ij}{n_\s} - f_5(e_\s) \right) e_\s \ ,
\llabel{090515c}
\end{eqnarray}
where 
$
f_4(e) = (1 + 3e^2/2 + e^4/8)/(1-e^2)^5 \ , \llabel{090515d}
$
$
f_5(e) = (1 + 15e^2/4 + 15e^4/8 + 5e^6/64)/(1-e^2)^{13/2} \ . \llabel{090515e}
$
For gaseous planets and rocky satellites we usually have $ k_{2_\p} \Delta t_\p \ll
k_{2_\s} \Delta t_\s $, and we can retain only terms in $K_\s$.

The ratio between orbital and spin evolution time-scales is roughly given by 
$ C_\s / (m_\s a_\s^2) \ll 1 $, meaning that the spin achieves an equilibrium position
($ \dot \vv{H}_\s = 0 $) much faster than the orbit.
Replacing the equilibrium rotation rate (Eq.\,\ref{090520a}) with $ \de_\s = 0 $
(for simplicity) in equations (\ref{090515b}) and (\ref{090515c}), gives:
\be
\dot a_\s = - \frac{7 K_\s}{m_\s a_\s} \, f_6(e_\s) e_\s^2
\ , \llabel{090522a}
\ee
\be
\dot e_\s = - \frac{7 K_\s}{2 m_\s a_\s^2} f_6(e_\s) (1-e_\s^2) e_\s 
\ , \llabel{090522b}
\ee
where 
$ 
f_6 (e) = (1 + 45e^2/14 + 8e^4 + 685e^6/224 + 255e^8/448 + 25e^{10}/1792)
(1-e^2)^{-15/2} / (1 + 3e^2 + 3e^4/8) \ . \llabel{090527a}
$
Thus, we always have $ \dot a_\s \le 0 $ and $ \dot e_\s \le 0 $, and the
final eccentricity is zero. 
Another consequence is that $ \dot \vv{L}_\s = - \dot \vv{H}_\p \simeq 0 $, and
the quantity $ a_\s (1 - e_\s^2) $ is conserved.
The final equilibrium semi-major axis is then given by
$
a_\es = a_\s (1 - e_\s^2) \ . \llabel{090522c}
$
However, from this point onwards, the tidal effects on the planet cannot be
neglected (Eq.\ref{090515b}), and they govern the future evolution of the satellite's
orbit.
For $ a_\es < a_\S $ or $ \de_\p \ge \pi / 2$, where $ a_\S^3 = G m_\p /
(\omega_\p \cos \de_\p)^2 $, the semi-major axis continues to decrease
until the satellite crashes into the planet, while in the remaining situations
it will increase.

\section{Application to Triton-Neptune}

Neptune's main satellite, Triton, presents unique features in the Solar System.
It is the only moon-sized body in a retrograde inclined orbit and the images
taken by the Voyager~2 spacecraft in 1989 revealed an extremely young surface
with few impact craters \citep[e.g.][]{Cruikshank_1995}.
This satellite should have remained molten until about 1\,Gyr ago and its
interior is still warm and geologically active considering its distance from the
Sun \citep{Schenk_Zahnle_2007}. 
Its composition also presents some similarities with Pluto
\citep{Tsurutani_etal_1990}.

These bizarre characteristics lead one to believe that Triton originally orbited
the Sun, belonging to the family of Kuiper-belt objects.
Most likely during the outward migration of Neptune, the orbits of the two
bodies intercepted and capture occurred.
This possibility is strongly supported by the fact that Triton's present orbit lies
between a group of small inner prograde satellites and a number of exterior
irregular satellites both prograde and retrograde.
Nereid, with an orbital eccentricity around 0.75, is also believed to have been
scattered from a regular satellite orbit 
\citep{McKinnon_1984}.

How exactly the capture occurred is still unknown, but some mechanisms have
been proposed: gas drag \citep{Pollack_etal_1979,McKinnon_Leith_1995},
a collision with a pre-existing regular satellite of Neptune
\citep{Goldreich_etal_1989}, or three-body interactions
\citep{Agnor_Hamilton_2006,Vokrouhlicky_etal_2008}.
All these scenarios require a very close passage to Neptune, and leave the planet
in eccentric orbits that must be damped by tides to the present one.
Tides are thus the only consensual mechanism acting on Triton's orbit.
The tidal distortion of Triton after a few close passages around Neptune, and the
consequent dissipation of tidal energy, can account for a substantial reduction
in the semi-major axis of its orbit, quickly bringing the planet from an
orbit outside Neptune's Hill sphere ($\sim$ 4700\,$R_\p$) to a bounded orbit.
Therefore, it cannot be ruled out that Triton was simply captured by tidal
interactions with Neptune after a close encounter in an almost parabolic orbit
\citep{McCord_1966}.

In this Letter we simulate the tidal evolution of the Triton-Neptune system
using the complete model described in Sect.2.
Triton is started in a very elliptical orbit with $e_\s=0.9968$ and a
semi-major axis of $a_\s = 2354$\,$R_\p $, corresponding to a final equilibrium
$ a_\es \simeq 15$\,$R_\p$, close to the present position of 
14.33\,$R_\p$. 
These specific values also place the satellite outside the Hill sphere of
Neptune at the apoapse and give a closest approach at a periapse of 7.53\,$R_\p$.
However, the non-secular perturbations of the Sun on Triton's orbit will cause
the eccentricity to vary around the mean value, allowing the periapse and apoapse
distances to attain lower and higher values \citep{Goldreich_etal_1989}.

For the radius of the bodies we use $ R_\p = 24\,764 $\,km and $ R_\s = 1\,353
$\,km \citep{Thomas_2000}, 
while for the masses, the $J_2$ of Neptune and the remaining orbital and spin
parameters we take the present values as determined by \citet{Jacobson_2009}.
For Triton we adopt $J_2 = 4.38 \times 10^{-4}$, the value measured for
Europa \citep{Anderson_etal_1998}, and $C_{22} = 0$, since our model does not
take into account spin-orbit resonances.
This choice is justified because Triton's observed topography never varies
beyond a kilometer \citep{Thomas_2000}. 
In addition, Triton should have undergone frequent collisions either with
other satellites of Neptune, or with external Kuiper-belt objects, and any capture
in a spin-orbit resonance different from the synchronous one, may not last for a
long time \citep{Stern_McKinnon_2000}. 
For tidal dissipation we adopt the same parameters as previous studies, that is,
$k_{2_\p} = 0.407 $ and $ Q_\p = 9000 $ \citep{Zhang_Hamilton_2008}, and 
$k_{2_\s} = 0.1 $ and $ Q_\s = 100 $
\citep{Goldreich_etal_1989,Chyba_etal_1989}, where $ Q_\ij^{-1} = \omega_\ij
\Delta t_\ij $. 

\begin{figure}
\centering
\includegraphics[width=8.9cm]{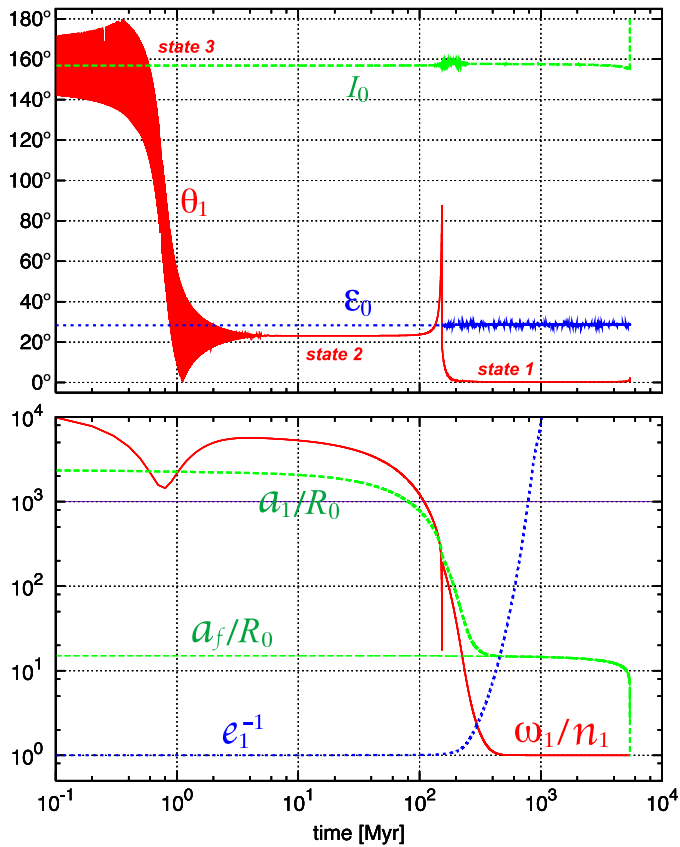}
\caption{Secular tidal evolution of the Triton-Neptune system. We plot the
orbital inclination to the Laplacian plane $I_\L$, the obliquity of Neptune to
the ecliptic $\ve_\p$ and the obliquity of Triton to the orbital plane $\de_\s$
(top), and the semi-major axis ratios $a_\s / R_\p $ and $a_\es / R_\p $, the
inverse of the eccentricity $e_\s^{-1}$ and Triton's rotation rate ratio
$\omega_\s / n_\s $ (bottom). 
\llabel{F1}}
\end{figure}

As for the orbit, the initial spin of Triton is unknown. We tested several
possibilities, but tides acting on the spin always drive it in the same way: the
rotation rate quickly evolves into the equilibrium value given by equation
(\ref{090520a}), while the obliquity is trapped in Cassini state 2, for $\de_\s
\simeq 180^\circ - I_\L = 23^\circ$ (Eq.\ref{090519e}).
In our standard simulation (Fig.\ref{F1}) we start Triton with a rotation period
of 24\,h and an obliquity $\de_\s = 170^\circ$ (librating around Cassini state
3).
In the very beginning, only Cassini states 2 and 3 exist ($\lambda_\s \ll 1 $),
but according to equation (\ref{090520e}),
equilibrium in Cassini state 3 cannot be maintained (Fig.\ref{F1}a).
As the semi-major axis decreases, the equilibrium obliquity for state 2
increases toward $90^\circ$ (Eq.\ref{090519e}).
At some point, the tidal torque becomes stronger than the conservative torque and
the spin quits this state.
The obliquity subsequently evolves into Cassini state 1 with $\de_\s
\simeq I_\L / \lambda_\s $ (Eq.\ref{090519e}).
The orbital inclination to the Laplacian plane is more or less constant, but it
presents some reduction when tides raised
by Triton on Neptune become important (present day situation).
At the very end of the evolution ($a_\s \simeq 7.79$\,$R_\p$), this trend is
reversed and the inclination quickly evolves into $180^\circ$. 
The obliquity of Neptune ($ \ve_\p $) does not undergo any significant
dissipation, but presents a secular oscillation of about one degree, from the
moment Triton becomes an inner satellite ($ a_\s < 100 $\,$R_\p$).
The semi-major axis and the eccentricity always decrease, as predicted by
equations (\ref{090522a}) and (\ref{090522b}), and the quantity $ a_\es =
a_\s (1-e_\s^2) $ is preserved, during the first stages of the evolution, with
the reduction observed being caused by tides on Neptune.
The eccentricity is very high during the first evolutionary stages, decreasing
very slowly.
However, as soon as Triton becomes an inner satellite, the eccentricity quickly
drops to a value very close to zero.
Finally, the rotation of the satellite decreases as the satellite orbit shrinks
into Neptune. It presents a rapid decrease when Cassini state 2 approaches
$90^\circ$, explained by Eq. (\ref{090520a}), and ultimately
stabilizes in the synchronous resonance, the presently observed configuration.

\section{Conclusion}

We have shown that the tidal evolution of a satellite can only be
correctly modeled when its spin is taken into account.
Previous studies adopted synchronous motion from the very beginning, which
is not the case for eccentric orbits. Tidal dissipation was therefore underestimated.
With the same tidal parameters used by \citet{Goldreich_etal_1989} we are able
to circularize Triton's orbit in only 1~Gyr.
Different tidal models and parameters may change the time-scale for the
evolution, but the global picture should remain the same.
We are also able to explain the present small value of Triton's
eccentricity $e_\s \simeq 10^{-5}$, as well as its small obliquity $\de_\s =
0.46^\circ$ and past evolution through Cassini states.
In particular, we can exclude the possibility that Triton was initially captured in state 1 and
determine when exactly the transition from state 2 to state 1 occurred
\citep{Chyba_etal_1989}.

Our study should also apply to the Moon, Charon and the satellites of Mars,
although in this case we need to take into account the quadropole moment of
inertia $C_{22} \neq 0$ \citep{Correia_2006}.
It can also be easily generalized to other stellar systems.


\vskip.5truecm
\hskip-.2truecm
The author thanks J.~Laskar 
for helpful discussions.
This work is dedicated to the memory of Ant\'onio and Nazareth Morgado.

\bibliographystyle{jphysicsB}
\bibliography{correia}

\end{document}